\documentclass[prl,twocolumn,reprint,superscriptaddress,showpacs,aps]{revtex4-1}

\usepackage{graphicx,amsmath,amssymb}
\usepackage{color}
\usepackage[normalem]{ulem}	
\usepackage{graphicx}

\begin{document}

\title{Simultaneous Occurrence of Multiferroism and the Short-Range Magnetic Order in DyFeO$_3$}

\author{Jinchen Wang}
\affiliation{Department of Physics, Renmin University of China, Beijing 100872, China}
\affiliation{Quantum Condensed Matter Division, Oak Ridge National Laboratory, Oak Ridge 37831, USA}
\affiliation{Department of Physics and Astronomy, University of Kentucky, Lexington 40506, USA}

\author{Juanjuan Liu}
\affiliation{Department of Physics, Renmin University of China, Beijing 100872, China}

\author{Jieming Sheng}
\author{Wei Luo}
\affiliation{Department of Physics, Renmin University of China, Beijing 100872, China}

\author{Feng Ye}
\affiliation{Quantum Condensed Matter Division, Oak Ridge National Laboratory, Oak Ridge 37831, USA}
\affiliation{Department of Physics and Astronomy, University of Kentucky, Lexington 40506, USA}

\author{Zhiying Zhao}

\author{Xuefeng Sun}
\affiliation{Hefei National Laboratory for Physical Sciences at Microscale, University of Science and Technology of China, Hefei, Anhui 230026, People's Republic of China}
\affiliation{Key Laboratory of Strongly-Coupled Quantum Matter Physics, Chinese Academy of Sciences, Hefei, Anhui 230026, People's Republic of China}
\affiliation{Collaborative Innovation Center of Advanced Microstructures, Nanjing, Jiangsu 210093, People's Republic of China}

\author{Sergey A. Danilkin}

\author{Guochu Deng}
\affiliation{Bragg Institute, ANSTO, Locked Bag 2001, Kirrawee DC NSW 2232, Australia}

\author{Wei Bao}
\email{wbao@ruc.edu.cn}
\affiliation{Department of Physics, Renmin University of China, Beijing 100872, China}

\begin{abstract}
We report a combined neutron scattering and magnetization study on the multiferroic DyFeO$_3$ which shows a very strong magnetoelectric effect. Applying magnetic field along the $c$-axis, the weak ferromagnetic order of the Fe ions is quickly recovered from a spin reorientation transition, and the long-range antiferromagnetic order of Dy becomes a short-range one. We found that the short-range order concurs with the multiferroic phase and is responsible for its sizable hysteresis. Our $H$-$T$ phase diagram suggests that the strong magnetoelectric effect in DyFeO$_3$ has to be understood with not only the weak ferromagnetism of Fe but also the short-range antiferromagnetic order of Dy.
\end{abstract}

\pacs{75.85.+t,75.25.-j,75.30.Kz}

\maketitle


Multiferroic (MF) materials, which possess two or more (anti-)ferroic orders, are topical for both fundamental interest and their application potentials \cite{nmat_6_13,nmat_6_21,J.Phys.D_38_R123,Khomskii20061,Tokura_RepProgPhys}. Since the discovery of spontaneous magnetoelectric (ME) coupling in TbMnO$_3$ \cite{kimura_2003}, a lot of materials have been actively explored in hope to realize a mutual control of magnetization or electric polarization using electric or magnetic field.
For Type II multiferroics, ferroelectricity has magnetic origin, and study of
rare-earth iron oxides $Re$FeO$_3$ have become a fertile ground for exploring such a multiferroic effect. 
DyFeO$_3$ has been discovered 
to have one of the strongest ME couplings \cite{PhysRevLett.101.097205} and more recently piezomagnetoelectric effect \cite{Nakajima2015}. Electric-field generation and reversal of ferromagnetic moments in a single-component bulk material have been realized in Tb and Gd doped DyFeO$_3$ \cite{Tokura2012_nphys838}. Multiferroic effects have since been observed also in many similar orthoferrites \cite{nmat_8_558,ApplPhysLett_102_062903,ApplPhysLett_105_052911}.

Research on rare-earth orthoferrites $Re$FeO$_3$ dates back to 1960s \cite{PhysRev_118_58,PhysRev.125.1843,JApplPhys_36_1033,JApplPhys_40_1061}. Taking advantage of strong coupling between the $Re^{3+}$ and $Fe^{3+}$ magnetic moments, these materials have shown a vast range of nontrivial magnetism, including solitonic lattice \cite{nmat_11_694}, laser induced spin switch \cite{Kimel_natphys_5_727,Kimel_nat_435_655}, and unconventional magnetization reversal \cite{CaoSX_Srep_4_5960,YuanSJ_PhysRevB.87.184405}.
For DyFeO$_3$, both rare-earth and iron moments develop complex magnetic orders at zero magnetic field. 
At $T^{Fe}_N\approx 600$ K, the Fe magnetic moments develop a canted antiferromagnetic order with a weak ferromagnetic (WFM) $c$-axis component due to the Dzyaloshinskii-Moriya interaction and it is denoted as the $G_xA_yF_z$ state in Bertaut's notation \cite{Bertaut}. The WFM state undergoes a spin reorientation through a Morin-type transition into the $A_xG_yC_z$ state at $T^{Fe}_R\approx 52$ K \cite{gorodetsky1968_jap1371}.
Below $T^{Dy}_N\approx 4$ K, magnetic moments of Dy also form a noncollinear antiferromagnetic (AFM) order in the $G_xA_y$ state \cite{JApplPhys_40_1061}. 
The multiferroic effect takes place when a magnetic field is applied in the $c$-axis \cite{PhysRevLett.101.097205}. 
The field recovers the WFM state and suppresses the Morin-type spin reorientation \cite{J.Phys.C_13_2567,Johnson1980557}. This recovery of WFM is currently believed to be crucial to the MF effect, from symmetry consideration as well as the proposed exchange striction mechanism \cite{PhysRevLett.101.097205,Tokura2012_nphys838,PhysRevB.8.5187,NewJ.Phys_12_093026}. The Dy AFM order is also one of the ingredients in the exchange striction mechanism, but is generally regarded as being field insensitive.
So far, no direct experimental evidence exists for the finite-field magnetic phases proposed in the theoretical mechanism.
Recent experiments suggested that more complex magnetic behaviors might occur in the low temperature regime where the MF effect appears \cite{Sun14_PhysRevB.89.224405}.

\begin{figure}[t!]
\includegraphics[width=1\columnwidth]{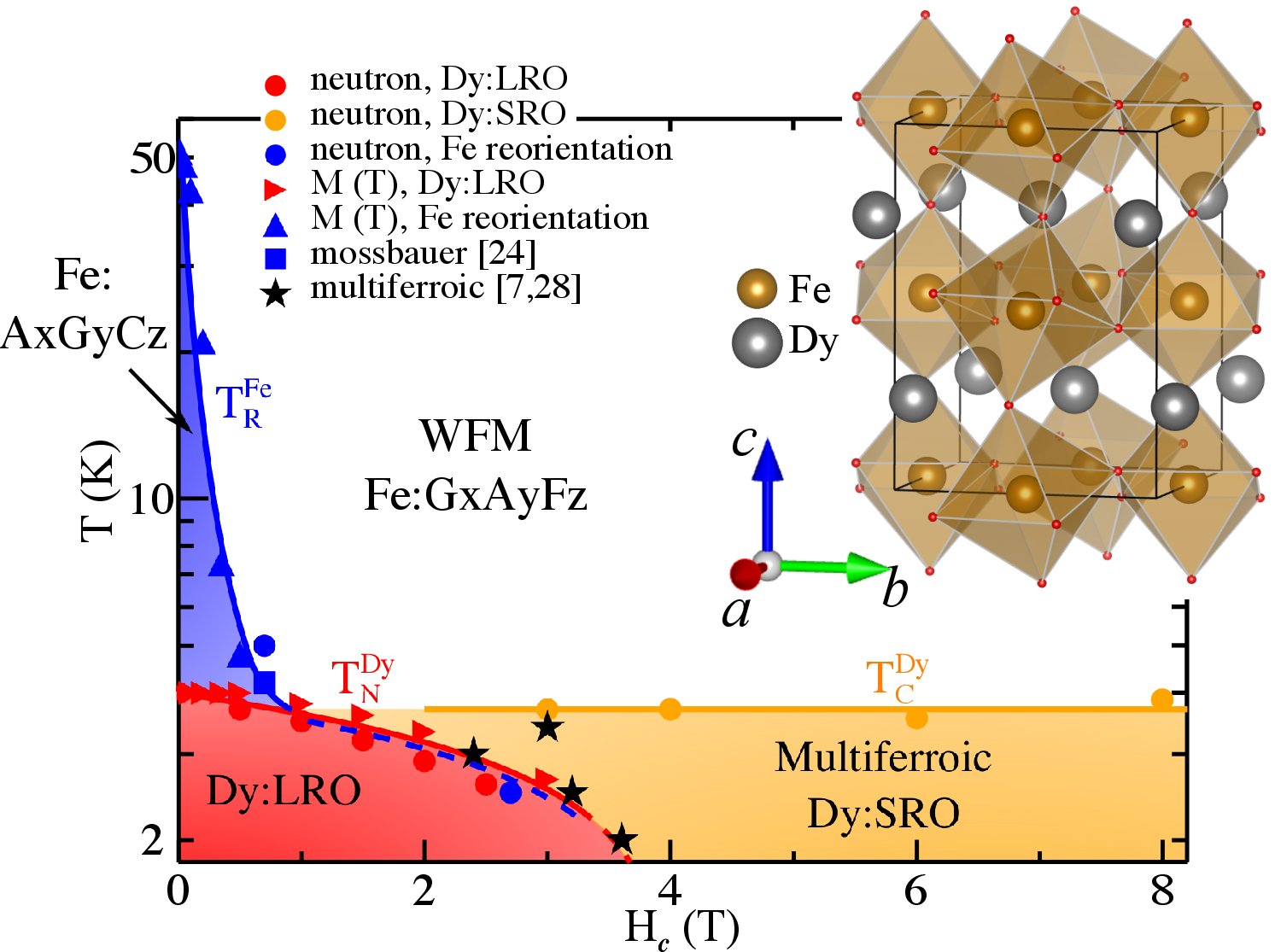}
\vskip -.2cm
\caption{The phase diagram of DyFeO$_3$ under the $c$-axis magnetic field. Logarithmic scale is used for temperature.
The red symbols denote Dy long range order temperatures $T_N^{Dy}$, circles from our neutron measurements and triangles from our magnetization data. 
The orange circles from our neutron data mark the Dy short range order temperatures $T_C^{Dy}$.
The blue symbols denote Fe spin reorientation transition temperatures $T_R^{Fe}$, with blue triangles from our magnetization measurements, blue circle from our neutron and squares from previous m$\ddot{o}$ssbauer data \cite{J.Phys.C_13_2567}.
Black stars are multiferroic transitions taken from literature \cite{PhysRevLett.101.097205,Sun14_PhysRevB.89.224405}.
Inset: Crystal structure of DyFeO$_3$ in an orthorhombic $Pbnm$ unit cell.
}
\label{pd}
\end{figure}

In this work, we establish the magnetic phase-diagram of DyFeO$_3$, see Fig.~\ref{pd}, through bulk magnetization and single-crystal neutron diffraction studies in magnetic field. The WFM phase occupies a large area next to $T^{Fe}_R$ in the phase diagram. Contrary to the current expectation, the AFM long-range order (LRO) of the Dy ions is suppressed by the magnetic field and it is replaced by a short-range order (SRO). The
strong multiferroic effect is observed only in the Dy SRO phase.
The concurrence of the SRO and MF phases suggests that the mechanism of this remarkably strong ME coupling in
the prototype rare-earth orthoferrite DyFeO$_3$ ought to be investigated through the interplay between the weak ferromagnetism of the Fe ions and the antiferromagnetic SRO of the Dy ions.


DyFeO$_3$ single crystals were grown by the floating-zone technique \cite{Sun14_PhysRevB.89.224405}.
It is crystallized into orthorhombic  unit cell with $a=5.261 \mathring{A}, b=5.489 \mathring{A}, c=7.542 \mathring{A}$ in the $Pbnm$ (No.~62) space group \cite{struc}.
The magnetization was measured using a Quantum Design VSM system with the applied magnetic field along the $c$-axis.
Neutron diffraction experiments were performed at the Taipan triple-axis spectrometer \cite{taipan} in Bragg Institute, Australia Nuclear Science and Technology Organization.
A Pyrolytic Graphite filter was put before the sample in the path of the 14.7 meV neutron beam, and a $40'$ collimator was used after the sample.
External fields along the $c$-axis were applied by an Oxford 12 T vertical field cryomagnet. 
Additional neutron diffraction experiments were conducted at the CORELLI Time-of-Flight spectrometer at Spallation Neutron Source (SNS), Oak Ridge National Laboratory, and the
temperature and magnetic field were regulated using a 5T vertical field cryomagnet.
During the experiments, magnetic fields were always changed at $T>T^{Fe}_R$ or $T_C^{Dy}$, and the constant-field temperature dependence studies follow a field cooling (FC) procedure.

\begin{figure}[thb!]
\includegraphics[width=0.95\columnwidth]{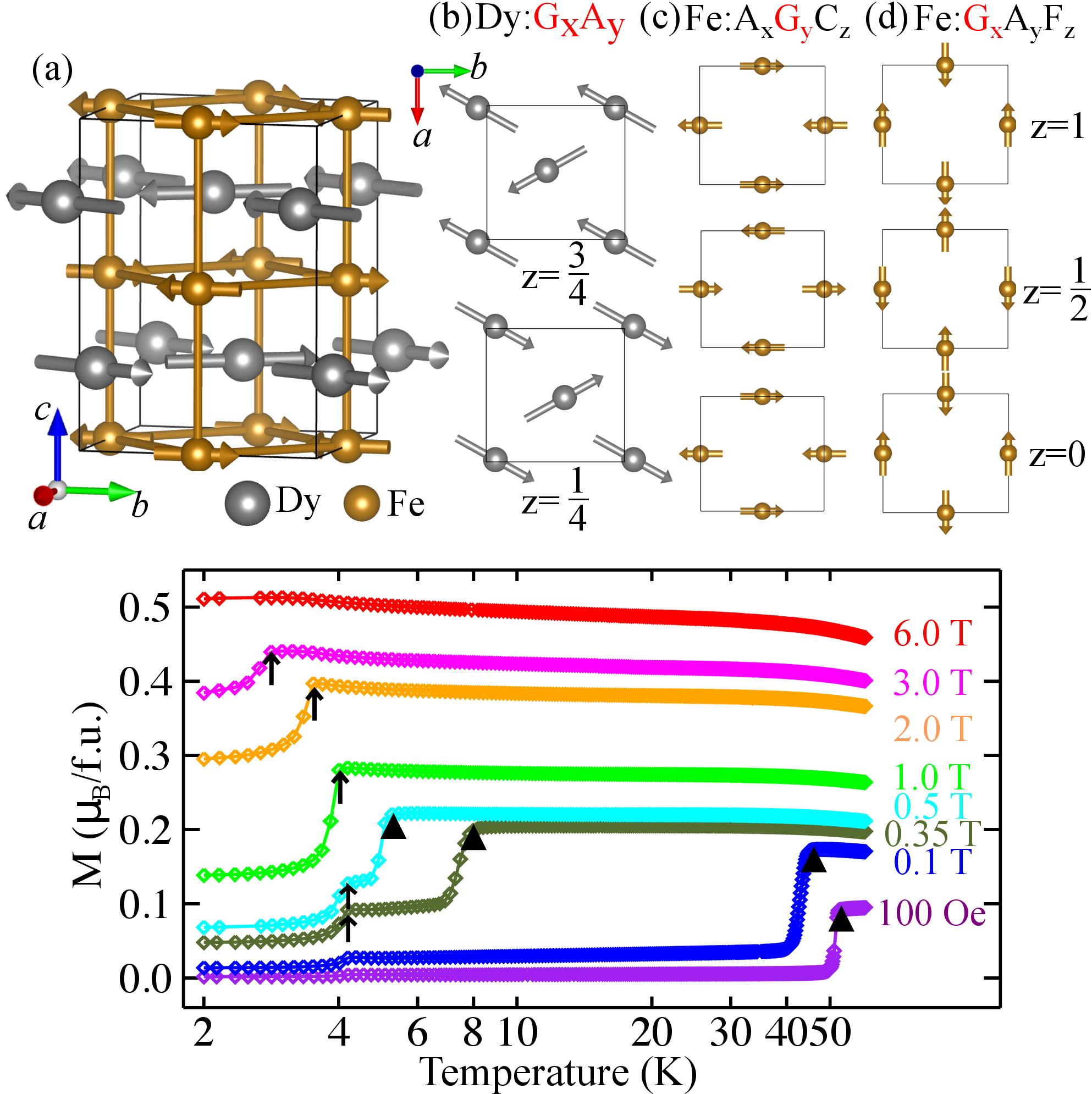}
\vskip -.2cm
\caption{
(a) The schematic magnetic structure of DyFeO$_3$ at zero field. 
(b) The in-plane view of the $G_xA_y$ configuration of Dy, and (c,d) the $A_xG_yC_z$ and $G_xA_yF_z$ configurations of Fe with a predominant $G$ component. The layers are denoted by the fractional coordinate $z$. 
(e) The magnetization data of DyFeO$_3$ under various $H_c$. Logarithmic scale is used for temperature.
The triangles mark the Fe spin reorientation temperature $T_R^{Fe}$, and arrows the AFM transition temperature $T_N^{Dy}$ of Dy.
The $H =$ 3 and 6 T data are shifted downwards by 0.07 and 0.32 $\mu_B/f.u.$ respectively.
}
\label{vsm}
\end{figure}

The determined phase diagram is shown in Fig.~\ref{pd}. The spin reorientation transition temperature $T_R^{Fe}$ separates two different magnetic structures of Fe, the WFM $G_xA_yF_z$ phase and the spin-reorientation $A_xG_yC_z$ phase. The reorientation temperature $T_R^{Fe}$ is suppressed quickly with magnetic field until the onset of the Dy order below $T_N^{Dy} <$ 4 K. 
Magnetic field below 4 K transforms the LRO of Dy into SRO while recovers the WFM of Fe at the same critical field.
At high fields, Dy ions only form short-range AFM order below $T_C^{Dy}$, in contrary to generally accepted assumption of a robust Dy AFM LRO in magnetic field \cite{PhysRevLett.101.097205,Tokura2012_nphys838,PhysRevB.8.5187,NewJ.Phys_12_093026}.
The multiferroic transitions, denoted by black stars, occur only in the coexistence phase of the WFM state of Fe and the SRO state of Dy marked by the orange color in Fig.~\ref{pd}. The detailed results will be present below.

\begin{figure}[tbh!]
\includegraphics[width=0.95\columnwidth]{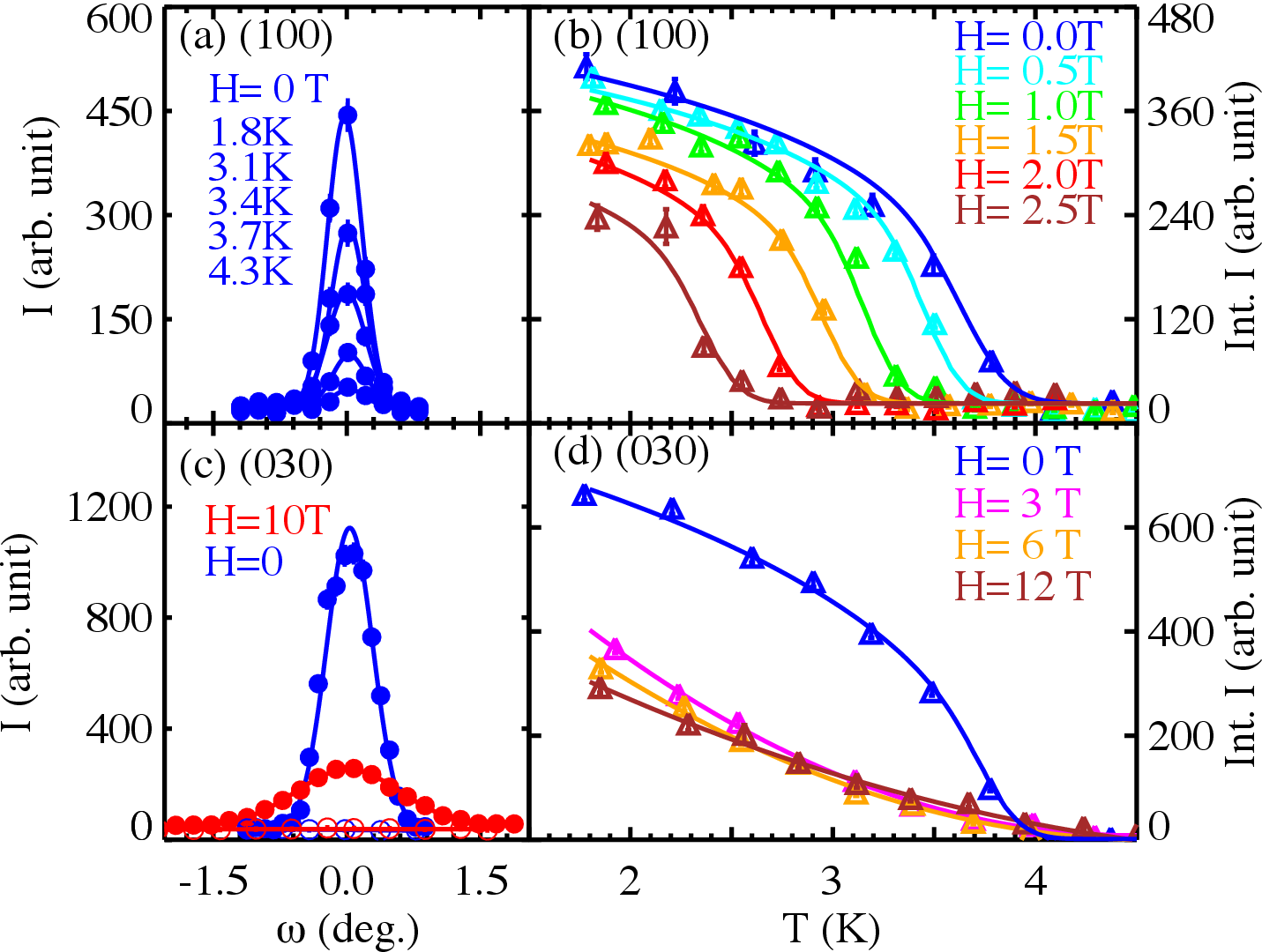}
\vskip -.2cm
\caption{	
	(a) The rocking scans of the Dy AFM Bragg peak (100) at selected temperatures, $H = 0$ T.
	(b) The integrated intensity of (100) as a function of temperature under various magnetic fields up to 2.5 T. 
	(c) The rocking scans of the Dy AFM Bragg peak (030) at 0 T and 10 T are compared. The solid circles were measured at 1.8 K, the open ones at 10 K.
	(d) The integrated intensity of (030) as a function of temperature at selected fields up to 12 T. 
}
\label{intn_t}
\end{figure}

The zero field magnetic structure is presented in Fig.~\ref{vsm} (a), with (b-d) presenting the in-plane view of various magnetic configurations. 
In Fig.~\ref{vsm} (e), the magnetization measurements clearly demonstrate two distinct transitions at the low fields.
Marked by black triangles, the Fe spin reorientation is observed by a significant decrease of magnetization due to the loss of WFM, consistent with earlier results \cite{gorodetsky1968_jap1371,Sun14_PhysRevB.89.224405}. Further cooling down, another drop of magnetization at around 4 K is observed, which corresponds to the ordering of Dy magnetic moments as will be shown by neutron diffraction data. 
By applying a small field, $T_R^{Fe}$ decreases rapidly, dropping from 52 K at 100 Oe to 5 K at 0.5 T. This can be understood as the field favors the WFM phase that contains the ferromagnetic component along the $c$ direction.
Under fields higher than 0.5 T, only the anomaly relating with the Dy AFM ordering can be resolved. 
It is getting weakened at higher fields in the magnitude of the anomaly and at the reduced transition temperature, consistent with the weaken Dy AFM LRO observed in neutron diffraction results to be shown in Fig.~\ref{intn_t} (b). 

Now let us turn to the neutron diffraction results. 
At zero field, both $(0kl)$ and $(hk0)$ scattering planes were investigated at Taipan. With vertical field applying in the $c$-axis, the horizontal $(hk0)$ is the scattering plane at Taipan and CORELLI spectrometers. But at CORELLI, a finite $\pm 5^o$ out of plane coverage can be achieved.
Above 4 K, magnetic Bragg scattering is solely from the contribution of Fe ions in the predominantly $G$-type AFM structure with wave vector (101) or (011) \cite{WFMneutron}. The $G_x$ to $G_y$ spin reorientation is confirmed by quantitative analysis of peak intensities that are changed around 52 K at $H_c$ = 0.  
The ordering of Dy moments is also confirmed below 4 K, signified by additional magnetic peaks that follow the intensity distribution of the $G_xA_y$ non-collinear AFM configuration \cite{Bidaux1968}. 
More detailed analysis is presented in Supplementary Material \cite{SM}.
To measure these different components, we benefit from the fact that the (100) or (300) Bragg peak contains only the $A_y$ component from Dy, the (010) or (030) only the $G_x$ from Dy, and the (200) or (020) from only nuclear contributions. 
Although the Fe magnetic structures contribute little Bragg scattering in the $(hk0)$ plane, it is possible to find (031) or (301) that is dominated by the Fe signals and within the reach of the experimental setup at CORELLI.

The temperature dependence of Dy magnetic peaks (100) and (030) at various $H_c$ was presented in Fig.~\ref{intn_t}.
Rather than being field insensitive as assumed previously, the AFM of Dy strongly depends on magnetic field. The onset temperature $T_N^{Dy}$ moves from 4 K to 2.6 K when $H_c$ ramps up from 0 T to 2.5 T, see Fig.~\ref{intn_t} (b). At higher magnetic fields, a completely different temperature dependence of the peak intensities appears, see Fig.~\ref{intn_t} (d).
The Bragg peaks are significantly broadened, in sharp contrast to the resolution limited peak profiles at zero field, as demonstrated in Fig.~\ref{intn_t} (c). The broadening of the Bragg peak and the concave shape of the order parameter at $H=3$-12 T in Fig.~\ref{intn_t} (d) indicate the high-field phase as a SRO. The broadening occurs on all peaks having Dy magnetic contributions, while pure nuclear peaks remain intact as demonstrated by Bragg peaks in the $(hk0)$ plane in Supplementary Material \cite{SM}.
The broadening could not be fitted by a small incommensurate peak splitting and is due to a reduced correlation length of the Dy AFM when the magnetic field is sufficiently strong.
Corrected for instrument resolution, the intrinsic widths measured on 15 different magnetic peaks fall into the range of $\sigma \sim 0.072 \pm 0.008 ~\mathring{A}^{-1}$, resulting in an estimated correlation length $\xi = 4\sqrt{\ln2}/\sigma \sim 46(5) ~\mathring{A}$ in the Dy SRO phase at 1.7 K and 4 T.
By comparing the integrated intensities of the same peaks in LRO phase at 0 T and SRO phase at 4 T, we noticed that the ratio of $I_{LRO} / I_{SRO}$ is about the same for all peaks we have measured. Thus the relative intensity distribution is not altered, and the same $G_xA_y$ correlation is likely maintained for the Dy ions in the SRO phase.
The onset temperature $T^{Dy}_C$ for SRO is about 4 K and shows little field dependence all the way up to 12 T, see Fig.~\ref{intn_t} (d). 

Magnetic moment of the Dy ion is as large as $gJ \sim 10 \mu_B$, and
the exchange energy between Dy ions scales with $k_B T_N^{Dy}$. At $H_c = 6$ T, the magnetic energy of full Dy polarization $gJH_c\sim 10k_B T_N^{Dy}$. From this point of view, it would be difficult for the Dy ions to maintain the AFM order at such a high field. Indeed, such magnetic polarized state have been observed in GdFeO$_3$ with the application of a field at about 3 T \cite{nmat_8_558}, close to the LRO breaking down field in our sample.
However, the ground state of Dy$^{3+}$ in DyFeO$_3$ is strongly Ising type as in the well studied DyAlO$_3$ \cite{gorodetsky1968_jap1371,Bidaux1968,Schuchert1969,Holmes1972} and the easy axis lies in the $ab$ plane 
\footnote{The local Ising direction changes between four sub-lattices following the symmetry of crystal structure, thus forming the in-plane non-collinear AFM structure of Dy.}. 
The magnetization at 6 T is still less than 1 $\mu_B/f.u.$ in both the paramagnetic and SRO states (see Fig.~\ref{vsm}), in contrast to the fully polarized 7 $\mu_B/f.u.$ in GdFeO$_3$. This may explain why the magnetic field is capable of breaking down the LRO but is not able to fully polarize the moment, eventually yielding a SRO. 
The competition between crystal field anisotropy, exchange interaction and Zeeman energy and their coupling to lattice degrees of freedom must be critical for the understanding of the experiment observations and the ME effect.
In any case, the break-down of LRO in DyFeO$_3$ observed in our experiments challenges the theory that the $G_xA_y$ AFM structure of Dy remains unchanged under applied magnetic field and the MF transition is a result of the Fe spin reorientation alone.

\begin{figure}[t!]
\includegraphics[width=0.95\columnwidth]{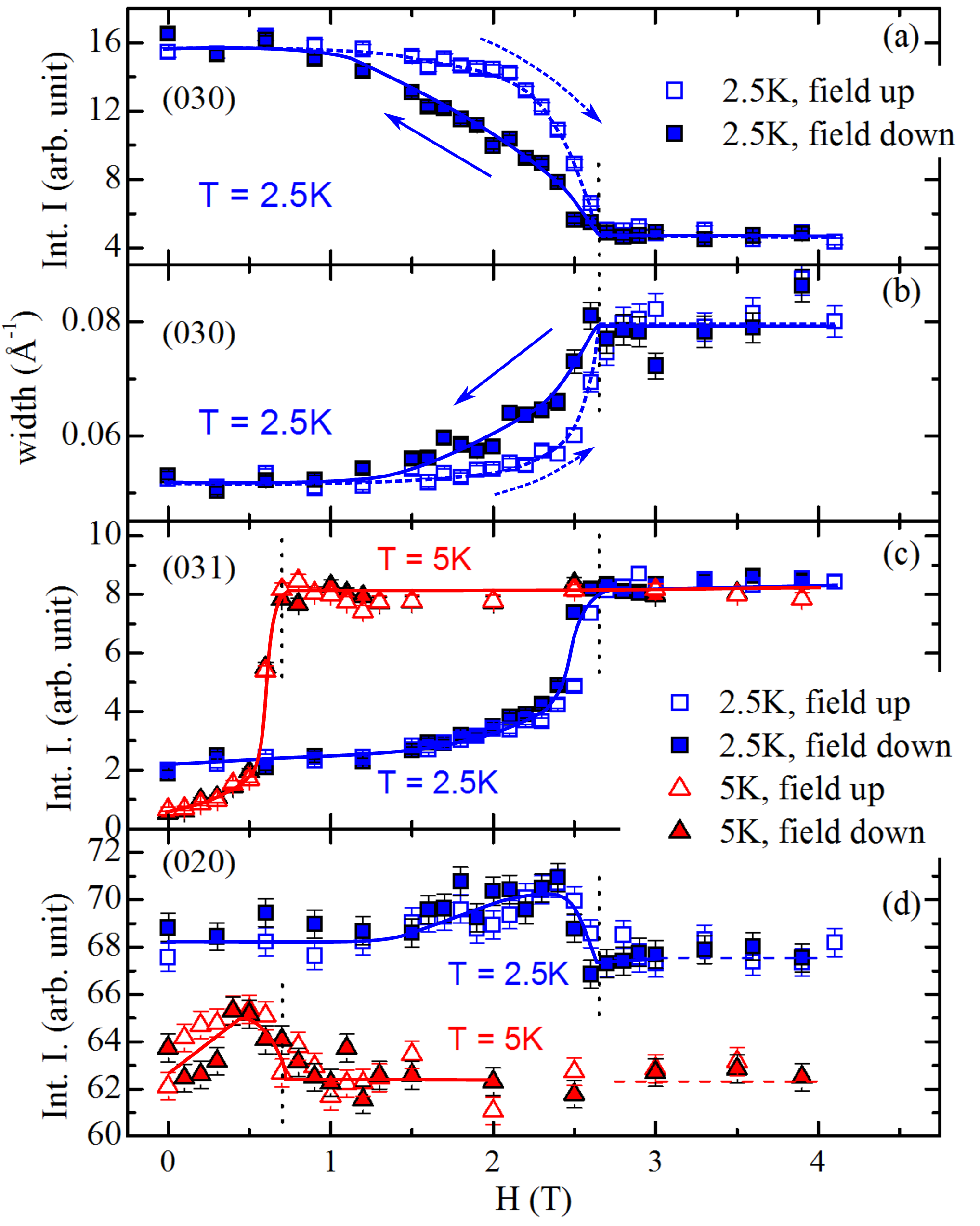}
\vskip -.2cm
\caption{The field dependence of the integrated intensity of (a) the (030), (c) the (031) and (d) the (020)
Bragg peak. (b) The peak width of the (030) Bragg peak represents the full width at half maximum (FWHM) in the radial direction. 
The field up and down processes are plotted with open and close symbols, respectively, at $T = 2.5$ K (blue) and 5  K (red). The critical fields are marked by the vertical dot lines. The 5 K curve in (d) is shifted downwards from the blue dash line for clarity.
}
\label{hdepd}
\end{figure}

In Fig.~\ref{hdepd}, we show field dependence of the (030) Bragg peak which is solely from the Dy ordering, the (020) solely from  nuclear origin, and the (031) that has mixed contributions from the Dy and Fe magnetic orders, but the Fe contribution gets much boosted after reorientation transition and dominates the intensity. The Dy contribution to (031) is weak in the LRO phase and will become weaker in the high field SRO phase. 
The field induced LRO to SRO transition is further demonstrated in Fig.~\ref{hdepd}(a,b). The magnetic field reduces the correlation length of the Dy AFM order as shown by the broaden peak width and reduces the Bragg intensity. The SRO stabilizes and becomes field independent above the critical field. The Fe reorientation occurs almost simultaneously with the Dy LRO to SRO transition, see Fig.~\ref{hdepd}(c). This explains why in magnetization measurement only one anomaly remains at high fields in Fig.~\ref{vsm} (e). The simultaneity also reveals a strong coupling between Dy and Fe magnetic moments, which is the key ingredient in designing functional magnetoelectric materials \cite{Tokura2012_nphys838}. 
The crystal structure also responds to the magnetic transition. At both 2.5 K and 5 K, a bump in the nuclear (020) intensity is observed at the critical field. It is interesting to note that such a magnetostrictive coupling exists at 5 K without the Dy magnetic ordering. But the multiferroice behavior emerges only when the Dy moments form short-range ordered AFM structure.

Significant hysteresis in DyFeO$_3$ was observed recently in magnetization, polarization and thermal conductivity measurements under magnetic field \cite{PhysRevLett.101.097205,Sun14_PhysRevB.89.224405}. A new phase of Fe magnetic order is proposed for the hysteresis region \cite{Sun14_PhysRevB.89.224405}. 
However, our neutron diffraction results in Fig.~\ref{hdepd} clearly show that there is no detectable hysteresis in the Fe magnetic transition. The hysteresis exists only in the Dy LRO-SRO transition. Thus, the sizable non-reversible effects in the bulk measurements \cite{Sun14_PhysRevB.89.224405} are due to the coupling to the Dy SRO.
Complex domain competition has been observed in other orthoferrites $Re$FeO$_3$ \cite{nmat_8_558}, and different tuning rate could lead to different macroscopic response \cite{Tokura2012_nphys838}. The hysteresis of the SRO could also play a role in these materials.

The SRO phase is important for the formation of electric polarization in the realization of the MF effect. 
If zero field cooled into well formed LRO phase, there's no polarization even under finite electric field in the paraelectric state (Fig.~4 in \cite{Sun14_PhysRevB.89.224405}), until close to the critical magnetic field where we now know that the Dy SRO state has appeared. 
According to the exchange striction mechanism, the spontaneous electric polarization emerges out of displacements of Dy$^{3+}$ ions \cite{PhysRevLett.101.097205}. Such displacement is likely disfavored by the LRO, and the relaxing into the SRO state could lower the energy. 
The poling treatment in multiferroic study usually helps to select one favorable domain. It is interesting that in the case of DyFeO$_3$, field cooling also helps to `freeze' the Dy moments into SRO state and hinders the restoring of LRO even when field is released. 

In summary, in the combined neutron diffraction and magnetization work on DyFeO$_3$, we found that the Fe spin reorientation phase is stabilized by the Dy long-range AFM order, both of which are suppressed by a magnetic field applying in the $c$-axis at $T_N^{Dy}(H_c)$. While the Fe order is replaced by the WFM order, the Dy LRO by the SRO of the same $G_xA_y$ AFM structure of a correlation length $\xi \sim 46(5) \mathring{A}$. The shortening of the correlation length and the break-up of the Dy LRO by the field are gradual and show substantial hysteresis. The appearance of the Dy SRO phase and its hysteresis coincide with the appearance and hysteresis of the magnetoelectric effect in this strong multiferroic material. 
The understanding of this prototype rare-earth orthoferrite multiferroic material ought to be refined by considering the
field-induced Dy antiferromagnetic short-range order in addition to the Fe WFM order.

The work at RUC and USTC was supported by National Basic Research Program of China (Grant Nos.~2012CB921700, 2011CBA00112 and 2015CB921201), the National Natural Science Foundation of China (Grant Nos.~11034012, 11190024, 11374277 and U1532147), and the Opening Project of Wuhan National High Magnetic Field Center (Grant No. PHMFF2015021). 
Research at SNS Oak Ridge National Laboratory was sponsored by the Scientific User Facilities Division, Office of Basic Energy Sciences, U.S. Department of Energy.
J.W.~acknowledges support from China Scholarship Council.

\section*{Supplementary Material}
\begin{figure}[thb!]
\includegraphics[width=0.7\columnwidth]{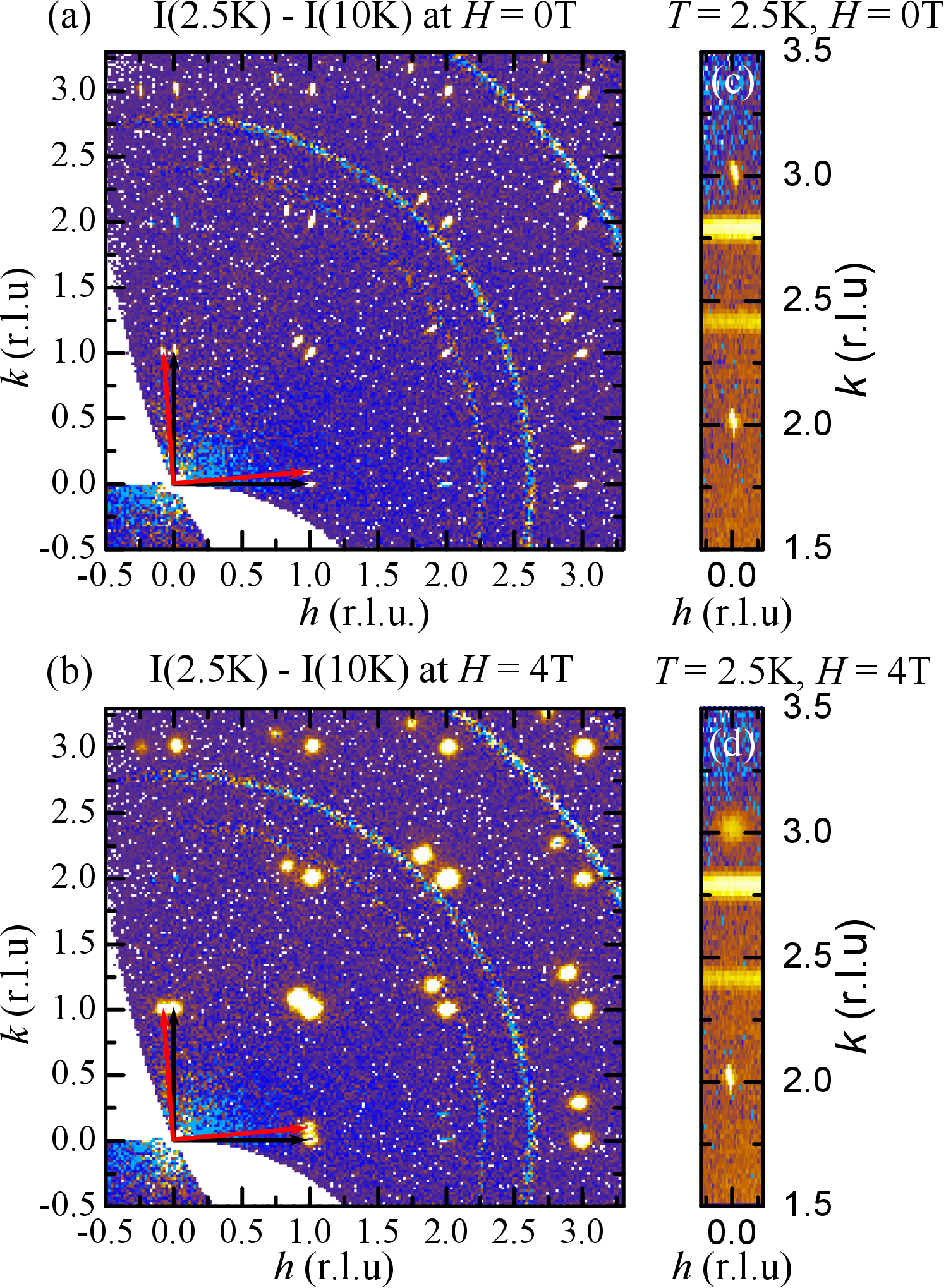}
\vskip -.2cm
\caption{Slice view of diffraction peaks in the $(hk0)$ scattering plane. In (a) 0 T and (b) 4 T, the backgrounds at 10 K are subtracted to highlight the Dy magnetic contributions. Two slightly rotated domains are noted, with black and red arrows denoting their coordinates. Raw data without the subtraction are shown in (c) 0 T and (d) 4 T, zoomed in the $-0.12 \le h \le 0.12$ region. The Aluminum powder rings come from sample environment such as the sample holder, cryogenics, etc.
}
\label{slice}
\end{figure}

Fig.~\ref{slice} presents the TOF neutron data in the $(hk0)$ scattering plane. By comparing the 2.5 K data at $H=0$ and 4 T in Fig.~\ref{slice}(a) and (b), which background at 10 K has been subtracted, it is evident that all  magnet peaks due to the Dy antiferromagnetic order are significantly broadened at the high field. Fig.~\ref{slice} (c) and (d) show raw data in the vicinity of the $h$ = 0 region without the subtraction. The pure nuclear peak (020) undergoes no change between $H=0$ and 4 T, while the pure Dy magnetic peak (030) gets considerably broadened. It demonstrates that the crystal structure remains the same, and only magnetic order becomes short-ranged.

When analyzing the neutron diffraction data, absorption is a significant issue since Dy is a strong neutron absorber. We cut the sample into a $5\times5.5\times1.25$ mm$^3$ slab to simplify the face indexed absorption correction. 
The direction cosines of different reflections were calculated, and the de Meulenaer \& Tompa algorithm \cite{Alcock1970} utilized by PLATON software \cite{PLATON} was used to perform numerical absorption correction. Data taken at Taipan with monochromatic beam was corrected and used for magnetic structure analysis. The correction for TOF data at CORELLI in semi-white beam is more challenging since different absorption coefficient has to be applied for peaks scattered from different wavelengths.

At $H=0$, diffraction data in the $(0kl)$ plane from Taipan was used to analysis the magnetic structure. 
Two-axis mode was used during data collection for the standard Lorentz correction \cite{tripleaxis}. 
We take advantage of the fact that in the $(0kl)$ plane, above $T_N^{Dy}$, the $G$-type Fe magnetic peaks at $k$ = odd and $l$ = even and the nuclear peaks at $k$ = even are separated in the reciprocal space. Below $T_N^{Dy}$, the $G$-type ($k$ = odd) and $A$-type ($k$ = even) magnetic peaks of Dy also have different Miller indeces and the magnetic intensity can be obtained by subtraction of the temperature independent nuclear intensity and the temperature insensitive Fe magnetic intensity above $T_N^{Dy}$. 
We collected Bragg peaks at 60 K for the $G_x$ type and at 10 K for the $G_y$ type of the Fe magnetic structure, respectively. For the $G_xA_y$ type of the Dy antiferromagnetic structure, the 1.5 K data were used after subtracting the 10 K data as background, given that the Debye-Waller factor is not significantly changed and Fe moment has already saturated in this temperature range.

From Fig.~\ref{slice}, two crystalline domains exist in the sample which are marked by the red and black coordinates and are rotated about the $c$-axis. Therefore all the $(00l)$ reflections were not used in refinement to ensure only one domain is contributed. The (010) and (011) 
were also abandoned since the Bragg peaks contain merged contributions from the two domains.

\begin{figure}[thb!]
	\includegraphics[width=0.95\columnwidth]{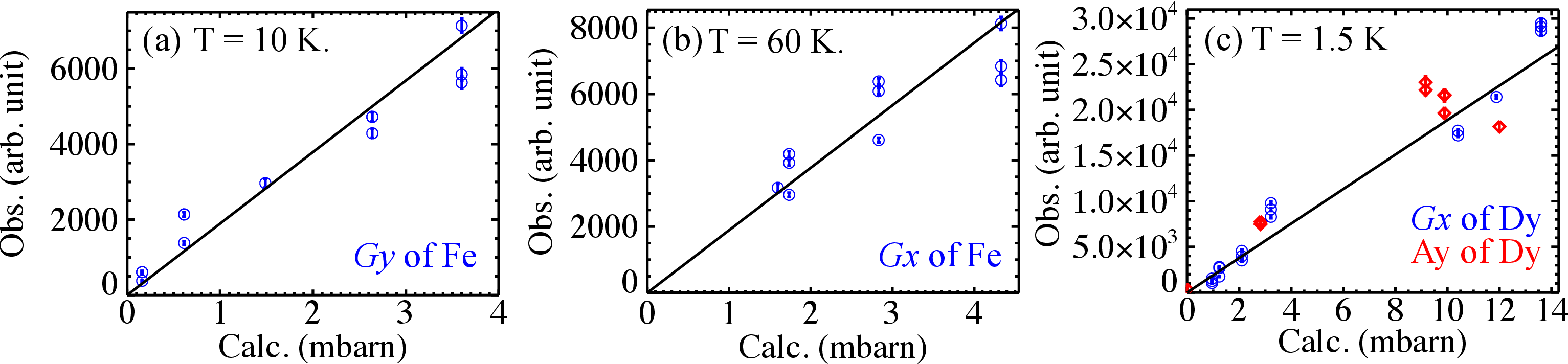}
	\vskip -.2cm
	\caption{The observed and calculated magnetic peak intensities for DyFeO$_3$ magnetic structure at $H=0$. 
		(a) $G_y$ peaks of Fe taken at 10 K.
		(b) $G_x$ peaks of Fe taken at 60 K.
		(c) $G_x$ (blue) and $A_y$ (red) peaks of Dy taken at 1.5 K.
	}
	\label{refine}
\end{figure}

The calculated peak intensities based on each magnetic configurations are compared with experimental results in Fig.~\ref{refine}.
The extinction effects must be severe for the nuclear peaks, most of which have large structure factor. Correction for the coupled extinction and absorption effect is challenging and we do not perform it. 
Thus we didn't use nuclear peaks to get the normalization scale factor for the magnetic moment calculation.
Instead, we used the previous magnetization results of $m_x=4.5 \mu_B$ for the $G_x$ and $m_y=8.0 \mu_B$ for the $A_y$ magnetic order \cite{gorodetsky1968_jap1371}, and compare the calculation with experiment data in Fig.~\ref{refine}(c). For the best convergence of the data on a straight line, the easy axis of the Dy moments is estimated as along the direction 30$^\circ$ from the $b$-axis in the $a$-$b$ plane. 
Using the $m_x$ and $m_y$ values of the Dy moment as the scale, the Fe moment can be refined to be $2.8 \mu_B$ at both 10 and 60 K from the $G_y$ and the $G_x$ data shown in Fig.~\ref{refine}(a) and (b), respectively.

\begin{figure}[thb!]
	\includegraphics[width=0.8\columnwidth]{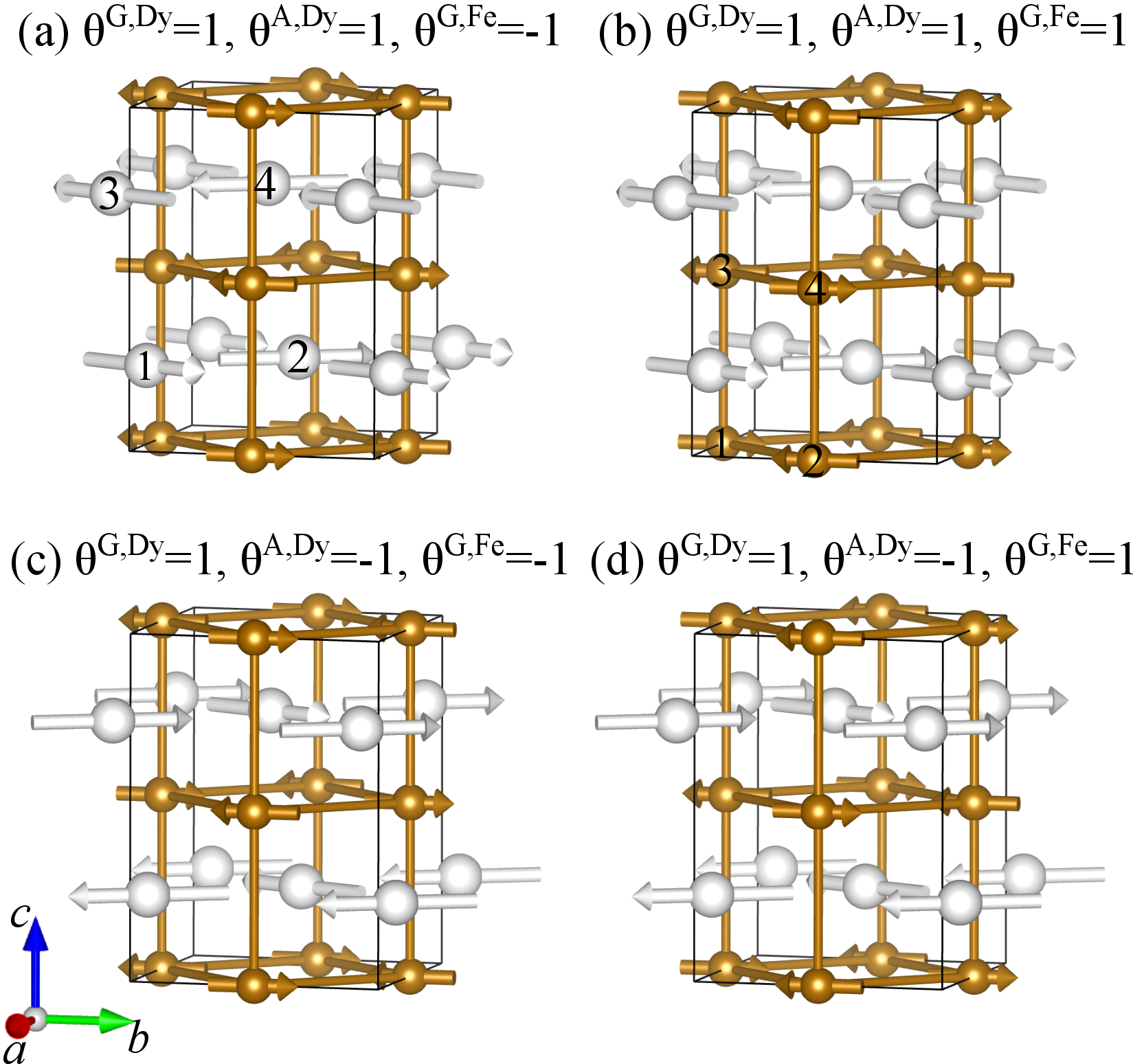}
	\vskip -.2cm
	\caption{The magnetic structures of different phase relations between the $G_x$ of Dy, the $A_y$ of Dy and the $G_y$ of Fe magnetic components.
		The four Fe and Dy ions in a unit cell are marked by the number.
	}
	\label{phase}
\end{figure}

It is worth mention that when both Fe and Dy order, there are phase issues between the $G$ type of the Fe magnetic order, the $A$ of Dy and the $G$ of the Dy magnetic orders.
Fig. \ref{phase} demonstrates the various combinations of the relative $\pi$-phase shifts among the $G$ phase of the Fe order and the $A$ and $G$ of the Dy orders.
The scattering intensity, rather than being the summation $|\vec{F}_{\bot}^{G,Fe}|^2 + |\vec{F}_{\bot}^{G,Dy}|^2 + |\vec{F}_{\bot}^{A,Dy}|^2$, should be proportional to
\begin{equation}
 |\sum_j f_j \vec{M}_{\bot,j} \exp{({i\vec{k}}\cdot\vec{r}_j)}|^2 = |\vec{F}_{\bot}^{G,Fe} + \vec{F}_{\bot}^{G,Dy} + \vec{F}_{\bot}^{A,Dy}|^2\nonumber
\label{sumf}
\end{equation}
where the subscript $j$ indexes the ironic site, $f$ the magnetic form factor, 
$\vec{M}$ and $\vec{M}_{\bot}$ the magnetic moment and its component perpendicular to the scattering vector ${\bf q}$,
$\vec{F}$ and $\vec{F}_{\bot}$ the scattering function for each magnetic order 
and its perpendicular component.
If we define the moment on the Fe1 ion at the $(0,\frac{1}{2},0)$ position as $\vec{M}({\rm Fe1}) = (0,\theta^{G,Fe} m^{Fe}_y,0)$, 
the moment on Dy1 at the $(x,y,\frac{1}{4})$ as $\vec{M}({\rm Dy1}) = (\theta^{G,Dy} m^{Dy}_x,\theta^{A,Dy} m^{Dy}_y,0)$, where $\theta=\pm 1$ denotes the relative phase factor,
the calculated scattering function will has the form $\vec{F}=(F_x^{G,Dy},F_y^{A,Dy}+F_y^{G,Fe},0)$, 
and $|\vec{F}_{\bot}|^2 = |\vec{F}|^2 - |\vec{F}\cdot \hat{\bf q}|^2$ would be
\begin{eqnarray}
|\vec{F}_{\bot}|^2 &= &|F_x^{G,Dy}|^2 + |F_y^{A,Dy}+F_y^{G,Fe}|^2 \nonumber \\ 
&&-|\hat{q}_xF_x^{G,Dy} + \hat{q}_y F_y^{A,Dy} + \hat{q}_yF_y^{G,Fe}|^2
\label{sumf2}
\end{eqnarray}
It can be verified that $F_y^{G,Fe}$ is real, $F_x^{G,Dy}$ and $\vec{F}_y^{A,Dy}$ are imaginary. As the result, the scattering function of Fe is decoupled from that of Dy in Eq.~\ref{sumf2}.
Therefore, the 10 K data which containing the Fe magnetic Bragg scattering can be used as the background for the Dy magnetic signal below $T_N^{Dy}$.
Additionally, 
the configurations in Fig.~\ref{phase} (a) and (b), or (c) and (d), are indistinguishable magnetic twins in our experiments.
On the other hand, the relative phase between $G$ and $A$ components of Dy does make a difference in peaks having both the $G$ and $A$ contributions. Therefore the magnetic structure in Fig.~\ref{phase}(a) is distinguishable from that in (c). Specifically, the Bragg peaks of (3,2,0) and (2,1,0) from the magnetic structure in Fig.~\ref{phase}(a) are
significant, while they are diminished from the magnetic structure in Fig.~\ref{phase} (c).
By examining the data in the $(hk0)$ plane, we concluded that the magnetic structure in Fig.~\ref{phase}(a) or (b) provides a better description of our data than that in (c) or (d). 
The data analysis for the Bragg peaks in the $(0,k,l)$ plane, shown in Fig.~\ref{refine}, takes advantage of the fact that the $G$ and $A$ components of the Dy antiferromagnetic structure contribute to Bragg peaks at different  positions. Thus, it is safe to analyse each of the $G$ of Fe, the $G$ of and the $A$ of Dy components independently.

\begin{thebibliography}{39}%
\makeatletter
\providecommand \@ifxundefined [1]{%
 \@ifx{#1\undefined}
}%
\providecommand \@ifnum [1]{%
 \ifnum #1\expandafter \@firstoftwo
 \else \expandafter \@secondoftwo
 \fi
}%
\providecommand \@ifx [1]{%
 \ifx #1\expandafter \@firstoftwo
 \else \expandafter \@secondoftwo
 \fi
}%
\providecommand \natexlab [1]{#1}%
\providecommand \enquote  [1]{``#1''}%
\providecommand \bibnamefont  [1]{#1}%
\providecommand \bibfnamefont [1]{#1}%
\providecommand \citenamefont [1]{#1}%
\providecommand \href@noop [0]{\@secondoftwo}%
\providecommand \href [0]{\begingroup \@sanitize@url \@href}%
\providecommand \@href[1]{\@@startlink{#1}\@@href}%
\providecommand \@@href[1]{\endgroup#1\@@endlink}%
\providecommand \@sanitize@url [0]{\catcode `\\12\catcode `\$12\catcode
  `\&12\catcode `\#12\catcode `\^12\catcode `\_12\catcode `\%12\relax}%
\providecommand \@@startlink[1]{}%
\providecommand \@@endlink[0]{}%
\providecommand \url  [0]{\begingroup\@sanitize@url \@url }%
\providecommand \@url [1]{\endgroup\@href {#1}{\urlprefix }}%
\providecommand \urlprefix  [0]{URL }%
\providecommand \Eprint [0]{\href }%
\providecommand \doibase [0]{http://dx.doi.org/}%
\providecommand \selectlanguage [0]{\@gobble}%
\providecommand \bibinfo  [0]{\@secondoftwo}%
\providecommand \bibfield  [0]{\@secondoftwo}%
\providecommand \translation [1]{[#1]}%
\providecommand \BibitemOpen [0]{}%
\providecommand \bibitemStop [0]{}%
\providecommand \bibitemNoStop [0]{.\EOS\space}%
\providecommand \EOS [0]{\spacefactor3000\relax}%
\providecommand \BibitemShut  [1]{\csname bibitem#1\endcsname}%
\let\auto@bib@innerbib\@empty
\bibitem [{\citenamefont {Cheong}\ and\ \citenamefont
  {Mostovoy}(2007)}]{nmat_6_13}%
  \BibitemOpen
  \bibfield  {author} {\bibinfo {author} {\bibfnamefont {S.-W.}\ \bibnamefont
  {Cheong}}\ and\ \bibinfo {author} {\bibfnamefont {M.}~\bibnamefont
  {Mostovoy}},\ }\href@noop {} {\bibfield  {journal} {\bibinfo  {journal}
  {Nature Mater.}\ }\textbf {\bibinfo {volume} {6}},\ \bibinfo {pages} {13}
  (\bibinfo {year} {2007})}\BibitemShut {NoStop}%
\bibitem [{\citenamefont {Ramesh}\ and\ \citenamefont
  {Spaldin}(2007)}]{nmat_6_21}%
  \BibitemOpen
  \bibfield  {author} {\bibinfo {author} {\bibfnamefont {R.}~\bibnamefont
  {Ramesh}}\ and\ \bibinfo {author} {\bibfnamefont {N.~A.}\ \bibnamefont
  {Spaldin}},\ }\href@noop {} {\bibfield  {journal} {\bibinfo  {journal}
  {Nature Mater.}\ }\textbf {\bibinfo {volume} {6}},\ \bibinfo {pages} {21}
  (\bibinfo {year} {2007})}\BibitemShut {NoStop}%
\bibitem [{\citenamefont {Fiebig}(2005)}]{J.Phys.D_38_R123}%
  \BibitemOpen
  \bibfield  {author} {\bibinfo {author} {\bibfnamefont {M.}~\bibnamefont
  {Fiebig}},\ }\href@noop {} {\bibfield  {journal} {\bibinfo  {journal}
  {Journal of Physics D: Applied Physics}\ }\textbf {\bibinfo {volume} {38}},\
  \bibinfo {pages} {R123} (\bibinfo {year} {2005})}\BibitemShut {NoStop}%
\bibitem [{\citenamefont {Khomskii}(2006)}]{Khomskii20061}%
  \BibitemOpen
  \bibfield  {author} {\bibinfo {author} {\bibfnamefont {D.}~\bibnamefont
  {Khomskii}},\ }\href@noop {} {\bibfield  {journal} {\bibinfo  {journal}
  {Journal of Magnetism and Magnetic Materials}\ }\textbf {\bibinfo {volume}
  {306}},\ \bibinfo {pages} {1 } (\bibinfo {year} {2006})}\BibitemShut
  {NoStop}%
\bibitem [{\citenamefont {Tokura}\ \emph {et~al.}(2014)\citenamefont {Tokura},
  \citenamefont {Seki},\ and\ \citenamefont {Nagaosa}}]{Tokura_RepProgPhys}%
  \BibitemOpen
  \bibfield  {author} {\bibinfo {author} {\bibfnamefont {Y.}~\bibnamefont
  {Tokura}}, \bibinfo {author} {\bibfnamefont {S.}~\bibnamefont {Seki}}, \ and\
  \bibinfo {author} {\bibfnamefont {N.}~\bibnamefont {Nagaosa}},\ }\href@noop
  {} {\bibfield  {journal} {\bibinfo  {journal} {Reports on Progress in
  Physics}\ }\textbf {\bibinfo {volume} {77}},\ \bibinfo {pages} {076501}
  (\bibinfo {year} {2014})}\BibitemShut {NoStop}%
\bibitem [{\citenamefont {Kimura}\ \emph {et~al.}(2003)\citenamefont {Kimura},
  \citenamefont {Goto}, \citenamefont {Shintani}, \citenamefont {Ishizaka},
  \citenamefont {Arima},\ and\ \citenamefont {Tokura}}]{kimura_2003}%
  \BibitemOpen
  \bibfield  {author} {\bibinfo {author} {\bibfnamefont {T.}~\bibnamefont
  {Kimura}}, \bibinfo {author} {\bibfnamefont {T.}~\bibnamefont {Goto}},
  \bibinfo {author} {\bibfnamefont {H.}~\bibnamefont {Shintani}}, \bibinfo
  {author} {\bibfnamefont {K.}~\bibnamefont {Ishizaka}}, \bibinfo {author}
  {\bibfnamefont {T.}~\bibnamefont {Arima}}, \ and\ \bibinfo {author}
  {\bibfnamefont {Y.}~\bibnamefont {Tokura}},\ }\href {\doibase
  10.1038/nature02018} {\bibfield  {journal} {\bibinfo  {journal} {Nature}\
  }\textbf {\bibinfo {volume} {426}},\ \bibinfo {pages} {55} (\bibinfo {year}
  {2003})}\BibitemShut {NoStop}%
\bibitem [{\citenamefont {Tokunaga}\ \emph {et~al.}(2008)\citenamefont
  {Tokunaga}, \citenamefont {Iguchi}, \citenamefont {Arima},\ and\
  \citenamefont {Tokura}}]{PhysRevLett.101.097205}%
  \BibitemOpen
  \bibfield  {author} {\bibinfo {author} {\bibfnamefont {Y.}~\bibnamefont
  {Tokunaga}}, \bibinfo {author} {\bibfnamefont {S.}~\bibnamefont {Iguchi}},
  \bibinfo {author} {\bibfnamefont {T.}~\bibnamefont {Arima}}, \ and\ \bibinfo
  {author} {\bibfnamefont {Y.}~\bibnamefont {Tokura}},\ }\href@noop {}
  {\bibfield  {journal} {\bibinfo  {journal} {Phys. Rev. Lett.}\ }\textbf
  {\bibinfo {volume} {101}},\ \bibinfo {pages} {097205} (\bibinfo {year}
  {2008})}\BibitemShut {NoStop}%
\bibitem [{\citenamefont {Nakajima}\ \emph {et~al.}(2015)\citenamefont
  {Nakajima}, \citenamefont {Tokunaga}, \citenamefont {Taguchi}, \citenamefont
  {Tokura},\ and\ \citenamefont {Arima}}]{Nakajima2015}%
  \BibitemOpen
  \bibfield  {author} {\bibinfo {author} {\bibfnamefont {T.}~\bibnamefont
  {Nakajima}}, \bibinfo {author} {\bibfnamefont {Y.}~\bibnamefont {Tokunaga}},
  \bibinfo {author} {\bibfnamefont {Y.}~\bibnamefont {Taguchi}}, \bibinfo
  {author} {\bibfnamefont {Y.}~\bibnamefont {Tokura}}, \ and\ \bibinfo {author}
  {\bibfnamefont {T.-h.}\ \bibnamefont {Arima}},\ }\href@noop {} {\bibfield
  {journal} {\bibinfo  {journal} {Phys. Rev. Lett.}\ }\textbf {\bibinfo
  {volume} {115}},\ \bibinfo {pages} {197205} (\bibinfo {year}
  {2015})}\BibitemShut {NoStop}%
\bibitem [{\citenamefont {Tokunaga}\ \emph {et~al.}(2012)\citenamefont
  {Tokunaga}, \citenamefont {Taguchi}, \citenamefont {Arima},\ and\
  \citenamefont {Tokura}}]{Tokura2012_nphys838}%
  \BibitemOpen
  \bibfield  {author} {\bibinfo {author} {\bibfnamefont {Y.}~\bibnamefont
  {Tokunaga}}, \bibinfo {author} {\bibfnamefont {Y.}~\bibnamefont {Taguchi}},
  \bibinfo {author} {\bibfnamefont {T.}~\bibnamefont {Arima}}, \ and\ \bibinfo
  {author} {\bibfnamefont {Y.}~\bibnamefont {Tokura}},\ }\href@noop {}
  {\bibfield  {journal} {\bibinfo  {journal} {Nature Physics}\ }\textbf
  {\bibinfo {volume} {8}},\ \bibinfo {pages} {838} (\bibinfo {year}
  {2012})}\BibitemShut {NoStop}%
\bibitem [{\citenamefont {Tokunaga}\ \emph {et~al.}(2009)\citenamefont
  {Tokunaga}, \citenamefont {Furukawa}, \citenamefont {Sakai}, \citenamefont
  {Taguchi}, \citenamefont {Arima},\ and\ \citenamefont {Tokura}}]{nmat_8_558}%
  \BibitemOpen
  \bibfield  {author} {\bibinfo {author} {\bibfnamefont {Y.}~\bibnamefont
  {Tokunaga}}, \bibinfo {author} {\bibfnamefont {N.}~\bibnamefont {Furukawa}},
  \bibinfo {author} {\bibfnamefont {H.}~\bibnamefont {Sakai}}, \bibinfo
  {author} {\bibfnamefont {Y.}~\bibnamefont {Taguchi}}, \bibinfo {author}
  {\bibfnamefont {T.}~\bibnamefont {Arima}}, \ and\ \bibinfo {author}
  {\bibfnamefont {Y.}~\bibnamefont {Tokura}},\ }\href@noop {} {\bibfield
  {journal} {\bibinfo  {journal} {Nature Mater.}\ }\textbf {\bibinfo {volume}
  {8}},\ \bibinfo {pages} {558} (\bibinfo {year} {2009})}\BibitemShut {NoStop}%
\bibitem [{\citenamefont {Shang}\ \emph {et~al.}(2013)\citenamefont {Shang},
  \citenamefont {Zhang}, \citenamefont {Zhang}, \citenamefont {Yuan},
  \citenamefont {Ge}, \citenamefont {Yuan},\ and\ \citenamefont
  {Feng}}]{ApplPhysLett_102_062903}%
  \BibitemOpen
  \bibfield  {author} {\bibinfo {author} {\bibfnamefont {M.}~\bibnamefont
  {Shang}}, \bibinfo {author} {\bibfnamefont {C.}~\bibnamefont {Zhang}},
  \bibinfo {author} {\bibfnamefont {T.}~\bibnamefont {Zhang}}, \bibinfo
  {author} {\bibfnamefont {L.}~\bibnamefont {Yuan}}, \bibinfo {author}
  {\bibfnamefont {L.}~\bibnamefont {Ge}}, \bibinfo {author} {\bibfnamefont
  {H.}~\bibnamefont {Yuan}}, \ and\ \bibinfo {author} {\bibfnamefont
  {S.}~\bibnamefont {Feng}},\ }\href@noop {} {\bibfield  {journal} {\bibinfo
  {journal} {Applied Physics Letters}\ }\textbf {\bibinfo {volume} {102}},\
  \bibinfo {pages} {062903} (\bibinfo {year} {2013})}\BibitemShut {NoStop}%
\bibitem [{\citenamefont {Chowdhury}\ \emph {et~al.}(2014)\citenamefont
  {Chowdhury}, \citenamefont {Goswami}, \citenamefont {Bhattacharya},
  \citenamefont {Ghosh}, \citenamefont {Basu},\ and\ \citenamefont
  {Neogi}}]{ApplPhysLett_105_052911}%
  \BibitemOpen
  \bibfield  {author} {\bibinfo {author} {\bibfnamefont {U.}~\bibnamefont
  {Chowdhury}}, \bibinfo {author} {\bibfnamefont {S.}~\bibnamefont {Goswami}},
  \bibinfo {author} {\bibfnamefont {D.}~\bibnamefont {Bhattacharya}}, \bibinfo
  {author} {\bibfnamefont {J.}~\bibnamefont {Ghosh}}, \bibinfo {author}
  {\bibfnamefont {S.}~\bibnamefont {Basu}}, \ and\ \bibinfo {author}
  {\bibfnamefont {S.}~\bibnamefont {Neogi}},\ }\href@noop {} {\bibfield
  {journal} {\bibinfo  {journal} {Applied Physics Letters}\ }\textbf {\bibinfo
  {volume} {105}},\ \bibinfo {pages} {052911} (\bibinfo {year}
  {2014})}\BibitemShut {NoStop}%
\bibitem [{\citenamefont {Koehler}\ \emph {et~al.}(1960)\citenamefont
  {Koehler}, \citenamefont {Wollan},\ and\ \citenamefont
  {Wilkinson}}]{PhysRev_118_58}%
  \BibitemOpen
  \bibfield  {author} {\bibinfo {author} {\bibfnamefont {W.~C.}\ \bibnamefont
  {Koehler}}, \bibinfo {author} {\bibfnamefont {E.~O.}\ \bibnamefont {Wollan}},
  \ and\ \bibinfo {author} {\bibfnamefont {M.~K.}\ \bibnamefont {Wilkinson}},\
  }\href@noop {} {\bibfield  {journal} {\bibinfo  {journal} {Phys. Rev.}\
  }\textbf {\bibinfo {volume} {118}},\ \bibinfo {pages} {58} (\bibinfo {year}
  {1960})}\BibitemShut {NoStop}%
\bibitem [{\citenamefont {Treves}(1962)}]{PhysRev.125.1843}%
  \BibitemOpen
  \bibfield  {author} {\bibinfo {author} {\bibfnamefont {D.}~\bibnamefont
  {Treves}},\ }\href@noop {} {\bibfield  {journal} {\bibinfo  {journal} {Phys.
  Rev.}\ }\textbf {\bibinfo {volume} {125}},\ \bibinfo {pages} {1843} (\bibinfo
  {year} {1962})}\BibitemShut {NoStop}%
\bibitem [{\citenamefont {Treves}(1965)}]{JApplPhys_36_1033}%
  \BibitemOpen
  \bibfield  {author} {\bibinfo {author} {\bibfnamefont {D.}~\bibnamefont
  {Treves}},\ }\href@noop {} {\bibfield  {journal} {\bibinfo  {journal}
  {Journal of Applied Physics}\ }\textbf {\bibinfo {volume} {36}},\ \bibinfo
  {pages} {1033} (\bibinfo {year} {1965})}\BibitemShut {NoStop}%
\bibitem [{\citenamefont {White}(1969)}]{JApplPhys_40_1061}%
  \BibitemOpen
  \bibfield  {author} {\bibinfo {author} {\bibfnamefont {R.~L.}\ \bibnamefont
  {White}},\ }\href@noop {} {\bibfield  {journal} {\bibinfo  {journal} {Journal
  of Applied Physics}\ }\textbf {\bibinfo {volume} {40}},\ \bibinfo {pages}
  {1061} (\bibinfo {year} {1969})}\BibitemShut {NoStop}%
\bibitem [{\citenamefont {Artyukhin}\ \emph {et~al.}(2012)\citenamefont
  {Artyukhin}, \citenamefont {Mostovoy}, \citenamefont {Jensen}, \citenamefont
  {Le}, \citenamefont {Prokes}, \citenamefont {de~Paula}, \citenamefont
  {Bordallo}, \citenamefont {Maljuk}, \citenamefont {Landsgesell},
  \citenamefont {Ryll}, \citenamefont {Klemke}, \citenamefont {Paeckel},
  \citenamefont {Kiefer}, \citenamefont {Lefmann}, \citenamefont {Kuhn},\ and\
  \citenamefont {Argyriou}}]{nmat_11_694}%
  \BibitemOpen
  \bibfield  {author} {\bibinfo {author} {\bibfnamefont {S.}~\bibnamefont
  {Artyukhin}}, \bibinfo {author} {\bibfnamefont {M.}~\bibnamefont {Mostovoy}},
  \bibinfo {author} {\bibfnamefont {N.~P.}\ \bibnamefont {Jensen}}, \bibinfo
  {author} {\bibfnamefont {D.}~\bibnamefont {Le}}, \bibinfo {author}
  {\bibfnamefont {K.}~\bibnamefont {Prokes}}, \bibinfo {author} {\bibfnamefont
  {V.~G.}\ \bibnamefont {de~Paula}}, \bibinfo {author} {\bibfnamefont {H.~N.}\
  \bibnamefont {Bordallo}}, \bibinfo {author} {\bibfnamefont {A.}~\bibnamefont
  {Maljuk}}, \bibinfo {author} {\bibfnamefont {S.}~\bibnamefont {Landsgesell}},
  \bibinfo {author} {\bibfnamefont {H.}~\bibnamefont {Ryll}}, \bibinfo {author}
  {\bibfnamefont {B.}~\bibnamefont {Klemke}}, \bibinfo {author} {\bibfnamefont
  {S.}~\bibnamefont {Paeckel}}, \bibinfo {author} {\bibfnamefont
  {K.}~\bibnamefont {Kiefer}}, \bibinfo {author} {\bibfnamefont
  {K.}~\bibnamefont {Lefmann}}, \bibinfo {author} {\bibfnamefont {L.~T.}\
  \bibnamefont {Kuhn}}, \ and\ \bibinfo {author} {\bibfnamefont {D.~N.}\
  \bibnamefont {Argyriou}},\ }\href@noop {} {\bibfield  {journal} {\bibinfo
  {journal} {Nature Materials}\ }\textbf {\bibinfo {volume} {11}},\ \bibinfo
  {pages} {694} (\bibinfo {year} {2012})}\BibitemShut {NoStop}%
\bibitem [{\citenamefont {Kimel}\ \emph {et~al.}(2009)\citenamefont {Kimel},
  \citenamefont {Ivanov}, \citenamefont {Pisarev}, \citenamefont {Usachev},
  \citenamefont {Kirilyuk},\ and\ \citenamefont
  {Rasing}}]{Kimel_natphys_5_727}%
  \BibitemOpen
  \bibfield  {author} {\bibinfo {author} {\bibfnamefont {A.~V.}\ \bibnamefont
  {Kimel}}, \bibinfo {author} {\bibfnamefont {B.~A.}\ \bibnamefont {Ivanov}},
  \bibinfo {author} {\bibfnamefont {R.~V.}\ \bibnamefont {Pisarev}}, \bibinfo
  {author} {\bibfnamefont {P.~A.}\ \bibnamefont {Usachev}}, \bibinfo {author}
  {\bibfnamefont {A.}~\bibnamefont {Kirilyuk}}, \ and\ \bibinfo {author}
  {\bibfnamefont {T.}~\bibnamefont {Rasing}},\ }\href@noop {} {\bibfield
  {journal} {\bibinfo  {journal} {Nature Physics}\ }\textbf {\bibinfo {volume}
  {5}},\ \bibinfo {pages} {727} (\bibinfo {year} {2009})}\BibitemShut {NoStop}%
\bibitem [{\citenamefont {Kimel}\ \emph {et~al.}(2005)\citenamefont {Kimel},
  \citenamefont {Kirilyuk}, \citenamefont {Usachev}, \citenamefont {Pisarev},
  \citenamefont {Balbashov3},\ and\ \citenamefont
  {Rasing}}]{Kimel_nat_435_655}%
  \BibitemOpen
  \bibfield  {author} {\bibinfo {author} {\bibfnamefont {A.~V.}\ \bibnamefont
  {Kimel}}, \bibinfo {author} {\bibfnamefont {A.}~\bibnamefont {Kirilyuk}},
  \bibinfo {author} {\bibfnamefont {P.~A.}\ \bibnamefont {Usachev}}, \bibinfo
  {author} {\bibfnamefont {R.~V.}\ \bibnamefont {Pisarev}}, \bibinfo {author}
  {\bibfnamefont {A.~M.}\ \bibnamefont {Balbashov3}}, \ and\ \bibinfo {author}
  {\bibfnamefont {T.}~\bibnamefont {Rasing}},\ }\href@noop {} {\bibfield
  {journal} {\bibinfo  {journal} {Nature}\ }\textbf {\bibinfo {volume} {435}},\
  \bibinfo {pages} {655} (\bibinfo {year} {2005})}\BibitemShut {NoStop}%
\bibitem [{\citenamefont {Cao}\ \emph {et~al.}(2014)\citenamefont {Cao},
  \citenamefont {Zhao}, \citenamefont {Kang}, \citenamefont {Zhang},\ and\
  \citenamefont {Ren}}]{CaoSX_Srep_4_5960}%
  \BibitemOpen
  \bibfield  {author} {\bibinfo {author} {\bibfnamefont {S.}~\bibnamefont
  {Cao}}, \bibinfo {author} {\bibfnamefont {H.}~\bibnamefont {Zhao}}, \bibinfo
  {author} {\bibfnamefont {B.}~\bibnamefont {Kang}}, \bibinfo {author}
  {\bibfnamefont {J.}~\bibnamefont {Zhang}}, \ and\ \bibinfo {author}
  {\bibfnamefont {W.}~\bibnamefont {Ren}},\ }\href@noop {} {\bibfield
  {journal} {\bibinfo  {journal} {Scientific Reports}\ }\textbf {\bibinfo
  {volume} {4}},\ \bibinfo {pages} {5960} (\bibinfo {year} {2014})}\BibitemShut
  {NoStop}%
\bibitem [{\citenamefont {Yuan}\ \emph {et~al.}(2013)\citenamefont {Yuan},
  \citenamefont {Ren}, \citenamefont {Hong}, \citenamefont {Wang},
  \citenamefont {Zhang}, \citenamefont {Bellaiche}, \citenamefont {Cao},\ and\
  \citenamefont {Cao}}]{YuanSJ_PhysRevB.87.184405}%
  \BibitemOpen
  \bibfield  {author} {\bibinfo {author} {\bibfnamefont {S.~J.}\ \bibnamefont
  {Yuan}}, \bibinfo {author} {\bibfnamefont {W.}~\bibnamefont {Ren}}, \bibinfo
  {author} {\bibfnamefont {F.}~\bibnamefont {Hong}}, \bibinfo {author}
  {\bibfnamefont {Y.~B.}\ \bibnamefont {Wang}}, \bibinfo {author}
  {\bibfnamefont {J.~C.}\ \bibnamefont {Zhang}}, \bibinfo {author}
  {\bibfnamefont {L.}~\bibnamefont {Bellaiche}}, \bibinfo {author}
  {\bibfnamefont {S.~X.}\ \bibnamefont {Cao}}, \ and\ \bibinfo {author}
  {\bibfnamefont {G.}~\bibnamefont {Cao}},\ }\href@noop {} {\bibfield
  {journal} {\bibinfo  {journal} {Phys. Rev. B}\ }\textbf {\bibinfo {volume}
  {87}},\ \bibinfo {pages} {184405} (\bibinfo {year} {2013})}\BibitemShut
  {NoStop}%
\bibitem [{\citenamefont {Bertaut}(1963)}]{Bertaut}%
  \BibitemOpen
  \bibfield  {author} {\bibinfo {author} {\bibfnamefont {E.~F.}\ \bibnamefont
  {Bertaut}},\ }\href@noop {} {\emph {\bibinfo {title} {Magnetism}}},\ edited
  by\ \bibinfo {editor} {\bibfnamefont {G.~T.}\ \bibnamefont {Rado}}\ and\
  \bibinfo {editor} {\bibfnamefont {H.}~\bibnamefont {Suhl}},\ Vol.~\bibinfo
  {volume} {3}\ (\bibinfo  {publisher} {Academic},\ \bibinfo {address} {New
  York.},\ \bibinfo {year} {1963})\BibitemShut {NoStop}%
\bibitem [{\citenamefont {Gorodetsky}\ \emph {et~al.}(1968)\citenamefont
  {Gorodetsky}, \citenamefont {Sharon},\ and\ \citenamefont
  {Shtrikman}}]{gorodetsky1968_jap1371}%
  \BibitemOpen
  \bibfield  {author} {\bibinfo {author} {\bibfnamefont {G.}~\bibnamefont
  {Gorodetsky}}, \bibinfo {author} {\bibfnamefont {B.}~\bibnamefont {Sharon}},
  \ and\ \bibinfo {author} {\bibfnamefont {S.}~\bibnamefont {Shtrikman}},\
  }\href@noop {} {\bibfield  {journal} {\bibinfo  {journal} {Journal of Applied
  Physics}\ }\textbf {\bibinfo {volume} {39}},\ \bibinfo {pages} {1371}
  (\bibinfo {year} {1968})}\BibitemShut {NoStop}%
\bibitem [{\citenamefont {Prelorendjo}\ \emph {et~al.}(1980)\citenamefont
  {Prelorendjo}, \citenamefont {Johnson}, \citenamefont {Thomas},\ and\
  \citenamefont {Wanklyn}}]{J.Phys.C_13_2567}%
  \BibitemOpen
  \bibfield  {author} {\bibinfo {author} {\bibfnamefont {L.~A.}\ \bibnamefont
  {Prelorendjo}}, \bibinfo {author} {\bibfnamefont {C.~E.}\ \bibnamefont
  {Johnson}}, \bibinfo {author} {\bibfnamefont {M.~F.}\ \bibnamefont {Thomas}},
  \ and\ \bibinfo {author} {\bibfnamefont {B.~M.}\ \bibnamefont {Wanklyn}},\
  }\href@noop {} {\bibfield  {journal} {\bibinfo  {journal} {Journal of Physics
  C: Solid State Physics}\ }\textbf {\bibinfo {volume} {13}},\ \bibinfo {pages}
  {2567} (\bibinfo {year} {1980})}\BibitemShut {NoStop}%
\bibitem [{\citenamefont {Johnson}\ \emph {et~al.}(1980)\citenamefont
  {Johnson}, \citenamefont {Prelorendjo},\ and\ \citenamefont
  {Thomas}}]{Johnson1980557}%
  \BibitemOpen
  \bibfield  {author} {\bibinfo {author} {\bibfnamefont {C.}~\bibnamefont
  {Johnson}}, \bibinfo {author} {\bibfnamefont {L.}~\bibnamefont
  {Prelorendjo}}, \ and\ \bibinfo {author} {\bibfnamefont {M.}~\bibnamefont
  {Thomas}},\ }\href@noop {} {\bibfield  {journal} {\bibinfo  {journal}
  {Journal of Magnetism and Magnetic Materials}\ }\textbf {\bibinfo {volume}
  {15 - 18, Part 2}},\ \bibinfo {pages} {557 } (\bibinfo {year}
  {1980})}\BibitemShut {NoStop}%
\bibitem [{\citenamefont {Yamaguchi}\ and\ \citenamefont
  {Tsushima}(1973)}]{PhysRevB.8.5187}%
  \BibitemOpen
  \bibfield  {author} {\bibinfo {author} {\bibfnamefont {T.}~\bibnamefont
  {Yamaguchi}}\ and\ \bibinfo {author} {\bibfnamefont {K.}~\bibnamefont
  {Tsushima}},\ }\href {\doibase 10.1103/PhysRevB.8.5187} {\bibfield  {journal}
  {\bibinfo  {journal} {Phys. Rev. B}\ }\textbf {\bibinfo {volume} {8}},\
  \bibinfo {pages} {5187} (\bibinfo {year} {1973})}\BibitemShut {NoStop}%
\bibitem [{\citenamefont {Stroppa}\ \emph {et~al.}(2010)\citenamefont
  {Stroppa}, \citenamefont {Marsman}, \citenamefont {Kresse},\ and\
  \citenamefont {Picozzi}}]{NewJ.Phys_12_093026}%
  \BibitemOpen
  \bibfield  {author} {\bibinfo {author} {\bibfnamefont {A.}~\bibnamefont
  {Stroppa}}, \bibinfo {author} {\bibfnamefont {M.}~\bibnamefont {Marsman}},
  \bibinfo {author} {\bibfnamefont {G.}~\bibnamefont {Kresse}}, \ and\ \bibinfo
  {author} {\bibfnamefont {S.}~\bibnamefont {Picozzi}},\ }\href@noop {}
  {\bibfield  {journal} {\bibinfo  {journal} {New Journal of Physics}\ }\textbf
  {\bibinfo {volume} {12}},\ \bibinfo {pages} {093026} (\bibinfo {year}
  {2010})}\BibitemShut {NoStop}%
\bibitem [{\citenamefont {Zhao}\ \emph {et~al.}(2014)\citenamefont {Zhao},
  \citenamefont {Zhao}, \citenamefont {Zhou}, \citenamefont {Zhang},
  \citenamefont {Li}, \citenamefont {Fan}, \citenamefont {Sun},\ and\
  \citenamefont {Li}}]{Sun14_PhysRevB.89.224405}%
  \BibitemOpen
  \bibfield  {author} {\bibinfo {author} {\bibfnamefont {Z.~Y.}\ \bibnamefont
  {Zhao}}, \bibinfo {author} {\bibfnamefont {X.}~\bibnamefont {Zhao}}, \bibinfo
  {author} {\bibfnamefont {H.~D.}\ \bibnamefont {Zhou}}, \bibinfo {author}
  {\bibfnamefont {F.~B.}\ \bibnamefont {Zhang}}, \bibinfo {author}
  {\bibfnamefont {Q.~J.}\ \bibnamefont {Li}}, \bibinfo {author} {\bibfnamefont
  {C.}~\bibnamefont {Fan}}, \bibinfo {author} {\bibfnamefont {X.~F.}\
  \bibnamefont {Sun}}, \ and\ \bibinfo {author} {\bibfnamefont {X.~G.}\
  \bibnamefont {Li}},\ }\href@noop {} {\bibfield  {journal} {\bibinfo
  {journal} {Phys. Rev. B}\ }\textbf {\bibinfo {volume} {89}},\ \bibinfo
  {pages} {224405} (\bibinfo {year} {2014})}\BibitemShut {NoStop}%
\bibitem [{\citenamefont {Marezio}\ \emph {et~al.}(1970)\citenamefont
  {Marezio}, \citenamefont {Remeika},\ and\ \citenamefont {Dernier}}]{struc}%
  \BibitemOpen
  \bibfield  {author} {\bibinfo {author} {\bibfnamefont {M.}~\bibnamefont
  {Marezio}}, \bibinfo {author} {\bibfnamefont {J.~P.}\ \bibnamefont
  {Remeika}}, \ and\ \bibinfo {author} {\bibfnamefont {P.~D.}\ \bibnamefont
  {Dernier}},\ }\href@noop {} {\bibfield  {journal} {\bibinfo  {journal} {Acta
  Crystallographica Section B}\ }\textbf {\bibinfo {volume} {26}},\ \bibinfo
  {pages} {2008} (\bibinfo {year} {1970})}\BibitemShut {NoStop}%
\bibitem [{\citenamefont {Danilkin}\ \emph {et~al.}(2007)\citenamefont
  {Danilkin}, \citenamefont {Horton}, \citenamefont {Moore}, \citenamefont
  {Braoudakis},\ and\ \citenamefont {Hagen}}]{taipan}%
  \BibitemOpen
  \bibfield  {author} {\bibinfo {author} {\bibfnamefont {S.~A.}\ \bibnamefont
  {Danilkin}}, \bibinfo {author} {\bibfnamefont {G.}~\bibnamefont {Horton}},
  \bibinfo {author} {\bibfnamefont {R.}~\bibnamefont {Moore}}, \bibinfo
  {author} {\bibfnamefont {G.}~\bibnamefont {Braoudakis}}, \ and\ \bibinfo
  {author} {\bibfnamefont {M.}~\bibnamefont {Hagen}},\ }\href {\doibase
  10.1080/10238160601045755} {\bibfield  {journal} {\bibinfo  {journal}
  {Journal of Neutron Research}\ }\textbf {\bibinfo {volume} {15}},\ \bibinfo
  {pages} {55} (\bibinfo {year} {2007})}\BibitemShut {NoStop}%
\bibitem [{WFM()}]{WFMneutron}%
  \BibitemOpen
  \href@noop {} {}\bibinfo {note} {The Dzyaloshinskii-Moriya interaction
  produces weak ferromagnetism. The weak ferromagnetic contribution of the
  canted AFM structure is not our focus in this study. We monitored the
  $G_xA_yF_z$ structure through its strongest $G_x$ component.}\BibitemShut
  {Stop}%
\bibitem [{\citenamefont {Bidaux}\ and\ \citenamefont
  {M{\'e}riel}(1968)}]{Bidaux1968}%
  \BibitemOpen
  \bibfield  {author} {\bibinfo {author} {\bibfnamefont {R.}~\bibnamefont
  {Bidaux}}\ and\ \bibinfo {author} {\bibfnamefont {P.}~\bibnamefont
  {M{\'e}riel}},\ }\href {\doibase 10.1051/jphys:01968002902-3022000}
  {\bibfield  {journal} {\bibinfo  {journal} {J. Phys. France}\ }\textbf
  {\bibinfo {volume} {29}},\ \bibinfo {pages} {220} (\bibinfo {year}
  {1968})}\BibitemShut {NoStop}%
\bibitem [{SM()}]{SM}%
  \BibitemOpen
  \href@noop {} {}\bibinfo {note} {See Supplementary Material}\BibitemShut
  {NoStop}%
\bibitem [{\citenamefont {Schuchert}\ \emph {et~al.}(1969)\citenamefont
  {Schuchert}, \citenamefont {H{\"u}fner},\ and\ \citenamefont
  {Faulhaber}}]{Schuchert1969}%
  \BibitemOpen
  \bibfield  {author} {\bibinfo {author} {\bibfnamefont {H.}~\bibnamefont
  {Schuchert}}, \bibinfo {author} {\bibfnamefont {S.}~\bibnamefont
  {H{\"u}fner}}, \ and\ \bibinfo {author} {\bibfnamefont {R.}~\bibnamefont
  {Faulhaber}},\ }\href {\doibase 10.1007/BF01392289} {\bibfield  {journal}
  {\bibinfo  {journal} {Z. Physik}\ }\textbf {\bibinfo {volume} {222}},\
  \bibinfo {pages} {105} (\bibinfo {year} {1969})}\BibitemShut {NoStop}%
\bibitem [{\citenamefont {Holmes}\ \emph {et~al.}(1972)\citenamefont {Holmes},
  \citenamefont {Van~Uitert}, \citenamefont {Hecker},\ and\ \citenamefont
  {Hull}}]{Holmes1972}%
  \BibitemOpen
  \bibfield  {author} {\bibinfo {author} {\bibfnamefont {L.~M.}\ \bibnamefont
  {Holmes}}, \bibinfo {author} {\bibfnamefont {L.~G.}\ \bibnamefont
  {Van~Uitert}}, \bibinfo {author} {\bibfnamefont {R.~R.}\ \bibnamefont
  {Hecker}}, \ and\ \bibinfo {author} {\bibfnamefont {G.~W.}\ \bibnamefont
  {Hull}},\ }\href {\doibase 10.1103/PhysRevB.5.138} {\bibfield  {journal}
  {\bibinfo  {journal} {Phys. Rev. B}\ }\textbf {\bibinfo {volume} {5}},\
  \bibinfo {pages} {138} (\bibinfo {year} {1972})}\BibitemShut {NoStop}%
\bibitem [{Note1()}]{Note1}%
  \BibitemOpen
  \bibinfo {note} {The local Ising direction changes between four sub-lattices
  following the symmetry of crystal structure, thus forming the in-plane
  non-collinear AFM structure of Dy.}\BibitemShut {Stop}%
\bibitem [{\citenamefont {Alcock}(1970)}]{Alcock1970}%
  \BibitemOpen
  \bibfield  {author} {\bibinfo {author} {\bibfnamefont {N.}~\bibnamefont
  {Alcock}},\ }\href@noop {} {\emph {\bibinfo {title} {Crystallographic
  Computing}}},\ edited by\ \bibinfo {editor} {\bibfnamefont {F.~R.}\
  \bibnamefont {Ahmed}}\ (\bibinfo  {publisher} {Munksgaard},\ \bibinfo
  {address} {Copenhagen},\ \bibinfo {year} {1970})\ pp.\ \bibinfo {pages}
  {271--276}\BibitemShut {NoStop}%
\bibitem [{\citenamefont {Spek}(2009)}]{PLATON}%
  \BibitemOpen
  \bibfield  {author} {\bibinfo {author} {\bibfnamefont {A.~L.}\ \bibnamefont
  {Spek}},\ }\href {\doibase 10.1107/S090744490804362X} {\bibfield  {journal}
  {\bibinfo  {journal} {Acta Cryst. D}\ }\textbf {\bibinfo {volume} {65}},\
  \bibinfo {pages} {148} (\bibinfo {year} {2009})}\BibitemShut {NoStop}%
\bibitem [{\citenamefont {Shirane}\ \emph {et~al.}(2002)\citenamefont
  {Shirane}, \citenamefont {Shapiro},\ and\ \citenamefont
  {Tranquada}}]{tripleaxis}%
  \BibitemOpen
  \bibfield  {author} {\bibinfo {author} {\bibfnamefont {G.}~\bibnamefont
  {Shirane}}, \bibinfo {author} {\bibfnamefont {S.~M.}\ \bibnamefont
  {Shapiro}}, \ and\ \bibinfo {author} {\bibfnamefont {J.~M.}\ \bibnamefont
  {Tranquada}},\ }\href@noop {} {\emph {\bibinfo {title} {Neutron scattering
  with a triple-axis spectrometer basic techniques}}}\ (\bibinfo  {publisher}
  {Cambridge University Press},\ \bibinfo {address} {Cambridge},\ \bibinfo
  {year} {2002})\BibitemShut {NoStop}%
\end{thebibliography}%

\end{document}